\documentclass[a4paper,11pt]{article}
\pdfoutput=1 

\usepackage{jinstpub} 
\usepackage{multirow}
\usepackage{subfigure}

\title{\boldmath Enhanced tunable cavity development for axion dark matter searches using a piezoelectric motor in combination with gears}

\author[a,1]{A. K. Yi,}
\author[b,1]{T. Seong,}\note{These authors contributed equally.}
\author[b]{S. Lee,}
\author[b]{S. Ahn,}
\author[b]{B. I. Ivanov,}
\author[b]{S. V. Uchaikin,}
\author[d,2]{B. R. Ko,\note{Corresponding author}}
\author[b,c]{and Y. K. Semertzidis}
\affiliation[a]{SLAC National Accelerator Laboratory, 2575 Sand Hill Rd., Menlo Park, California 94025, USA}
\affiliation[b]{Center for Axion and Precision Physics Research (CAPP), Institute for Basic Science (IBS), Daejeon 34051, Republic of Korea}
\affiliation[c]{Department of Physics, Korea Advanced Institute of Science and Technology (KAIST), Daejeon 34141, Republic of Korea}
\affiliation[d]{Department of Accelerator Science, Korea University, Sejong 30019, Republic of Korea}
\emailAdd{brko@korea.ac.kr}

\abstract{
  Most search experiments sensitive to quantum chromodynamics (QCD)
  axion dark matter benefit from microwave cavities, as
  electromagnetic resonators, that enhance the detectable axion signal
  power and thus the experimental sensitivity drastically.  
  As the possible axion mass spans multiple orders of magnitude,
  microwave cavities must be tunable and it is desirable for the
  cavity to have a tunable frequency range that is as wide as
  possible.  
  Since the tunable frequency range generally increases as the
  dimension of the conductor tuning rod increases for a given
  cylindrical conductor cavity system, we developed a cavity system
  with a large  dimensional tuning rod in order to increase this.
  We, for the first time, employed not only a piezoelectric motor, but
  also gears to drive a large and accordingly heavy tuning rod, where
  such a combination to increase driving power can be adopted for
  extreme environments as is the case for axion dark matter
  experiments: cryogenic, high-magnetic-field, and high vacuum.  
  Thanks to such higher power derived from the piezoelectric motor and
  gear combination, we realized a wideband tunable cavity whose
  frequency range is about 42\% of the central resonant frequency of
  the cavity, without sacrificing the experimental sensitivity too
  much.
}  

\begin{document}
\maketitle
\flushbottom

\section{Introduction}\label{sec:intro}
The QCD axion (or just axion)~\cite{AXION1,AXION2} is a very natural
solution for the strong $CP$ problem in the Standard Model of particle
physics (SM)~\cite{strongCP1,strongCP2,strongCP3,strongCP4,strongCP5}
proposed by Peccei and Quinn~\cite{PQ}, and is predicted to be
massive, stable, cold, and weakly interacting with the
SM~\cite{CDM_LOW1,CDM_LOW2,CDM_LOW3}. Such axion characteristics meet
those of cold dark matter (CDM) which is believed to constitute about
85\% of the matter in our Universe~\cite{PLANCK}. The QCD axion is one
of the leading CDM candidates and in this context is referred to as
axion dark matter and the axion haloscope
search~\cite{sikivie1,sikivie2} is the only search method sensitive to
axion dark matter to date thanks to resonant conversion from axions to
photons with help from a resonant cavity when the axion mass $m_a$
matches the cavity mode frequency $\nu$, $m_a=h\nu/c^2$.

As the unknown axion mass and resonant conversion by the microwave
cavity are the most important factors for this search, the figure of
merit in axion haloscope searches is the scanning rate~\cite{scanrate}
proportional to $B^4V^2C_{\rm mode}^2Q_{L_{\rm mode}}/T^2_n$, where
$B$ is the magnetic field inside the cavity volume, $V$ is the cavity
volume, $C_{\rm mode}$ is the cavity mode-dependent form
factor~\cite{EMFF_BRKO}, $Q_{L_{\rm mode}}$ is the loaded quality
factor of the cavity mode, and $T_n$ is the system noise temperature.
Among those parameters, the cavity is related to the parameters $V$,
$C_{\rm mode}$, and $Q_{L_{\rm mode}}$.
The cavity geometry is typically cylindrical for axion haloscope
searches using solenoid magnets and the employed cavity modes have
an electromagnetic field profile similar to that of the TM$_{010}$
mode of a cylinder to maximize the $C_{\rm mode}$. 
Frequency tuning is generally performed by moving the tuning rod such
that its position moves in the radial
direction~\cite{CAVITY_RSI,ADMX_NIMA,HAYSTAC_CAVITY}; rarely is it
ever tuned vertically~\cite{BNL1,BNL2}.
The power dissipated by the driving of the tuning rod could increase
the physical temperature of the cavity, and subsequently the $T_n$,
implying that the cavity could affect $T_n$ as well.

The axion parameter space that the axion haloscopes are probing is
enormously wide even if it is limited to the microwave region in light
of the best scanning rate to date~\cite{12TB_PRL} and axion haloscopes
can usually expand the parameter space by using several cavity systems
with different dimensions of the cavity walls or tuning rods. However,
such cavity production itself is by no means trivial.
Therefore, it is highly desirable for the cavities to
have tunable frequency ranges as wide as possible, as long as the
experimental sensitivity is retained. For a cylindrical conductor
cavity with a cylindrical conductor tuning rod, the cavity mode
frequencies and their tuning range are generally proportional to the
diameter of the tuning rod for a given cavity diameter. According to
the cavity simulations~\cite{CST,COMSOL} for the aforementioned
frequency tuning~\cite{CAVITY_RSI,ADMX_NIMA,HAYSTAC_CAVITY}, the
tunable range of a cavity system with a tuning rod dimension of about
a tenth of the cavity barrel is at best 30\% with respect to the
central frequency of the tuning range.

In this work, we developed a cavity system with a large, and therefore
heavy, tuning rod to increase the frequency tuning range. With a
piezoelectric motor in combination with gears, we were able to drive a
heavy tuning rod and realized a wideband tunable cavity whose
frequency range is about 42\% with respect to the central frequency of
the tuning range without compromising the experimental sensitivity too
much. Since the cavity mode frequencies increase accordingly
with the dimension of the conductor tuning rod, the increase from
30 to 42\% actually corresponds to increasing the absolute frequency
range by a factor of about 1.8, which is nontrivial.
Note that this is the first time that a piezoelectric motor and gear
combination has been employed in order to drive stronger power in the
extreme environment associated to axion dark matter experiments, i.e.,
cryogenic, high-magnetic-field, and high vacuum.
\section{Cavity system}\label{sec:cavity}
\begin{figure}
  \centering
  \includegraphics[width=0.9\textwidth]{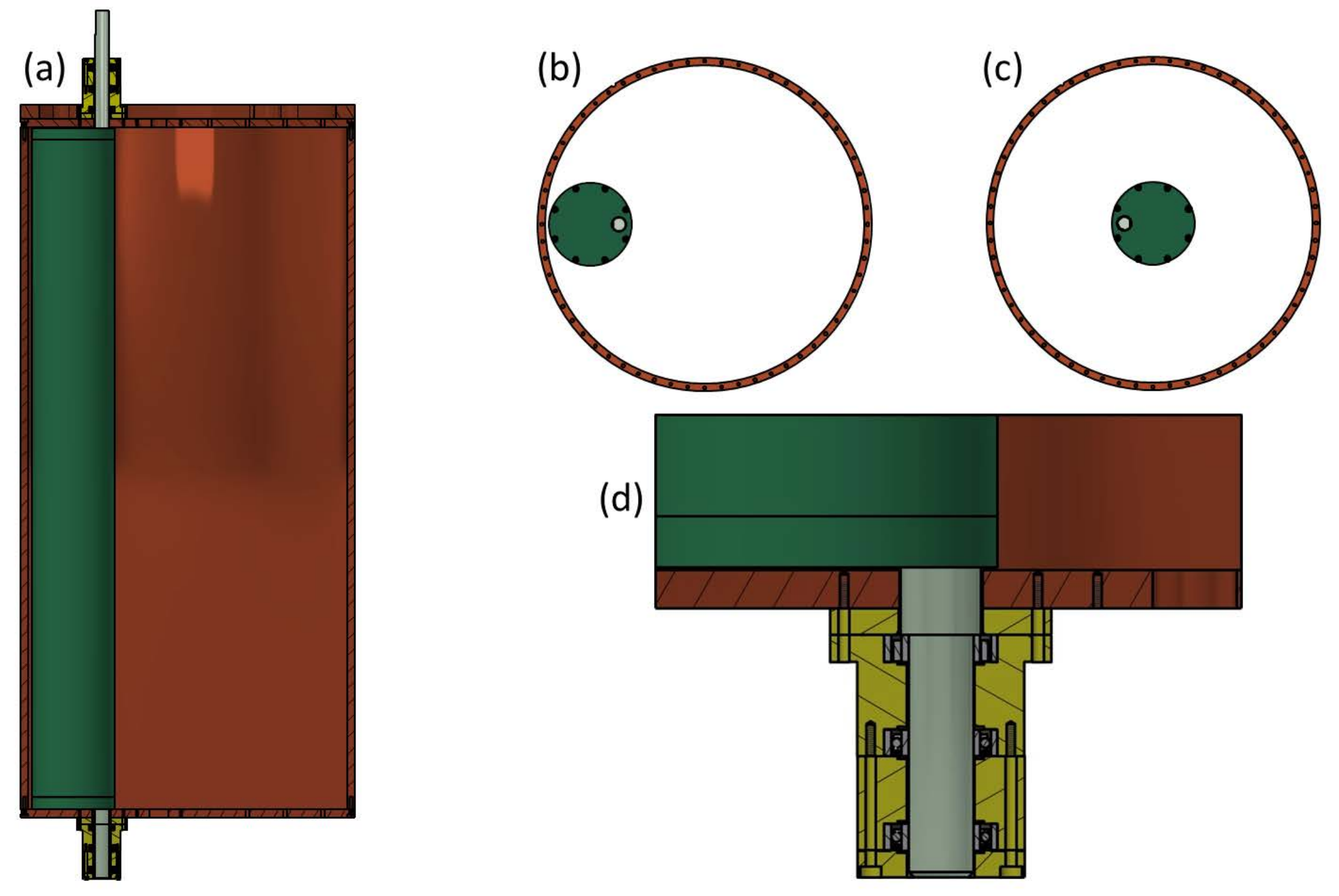}
  \caption{Overall of the cavity system. (a) is a lateral cutaway
    view showing the green colored tuning rod.
    (b) and (c) are top views with the end caps removed, where they
    show the tuning rod and axle locations for tuning the lowest and
    highest resonance frequencies, respectively.
    Note the grey colored alumina axle and eight OFHC copper screws on
    the end cap of the tuning rod are shown in (b) and (c) together
    with many screw holes in the cavity barrel. As one can see in (a),
    the yellow turrets with dark grey colored ceramic ball bearings
    mounted on the end caps align the axle.    
    (d) is a zoomed view around the bottom turret
    where the three ball bearings are in there and the uppermost
    bearing supports the tuning rod.    
    Albeit different colors, they are all OFHC copper except for the
    alumina axle with grey color.}  
  \label{FIG:CAVITY}
\end{figure}
The cavity system is constituted by a cylinder and a cylindrical
tuning rod, where the inner diameter of the cylinder is 262~mm and the
outer diameter of the tuning rod is 68~mm. The heights of the cylinder
and the tuning rod are 560 and 559~mm, respectively, and thus the gaps
between the end caps of the cavity and the tuning rod are
0.5~mm. Figure~\ref{FIG:CAVITY} shows the lateral and top views of the
cavity system.
\begin{figure}[h]
  \centering
  \includegraphics[width=0.7\textwidth]{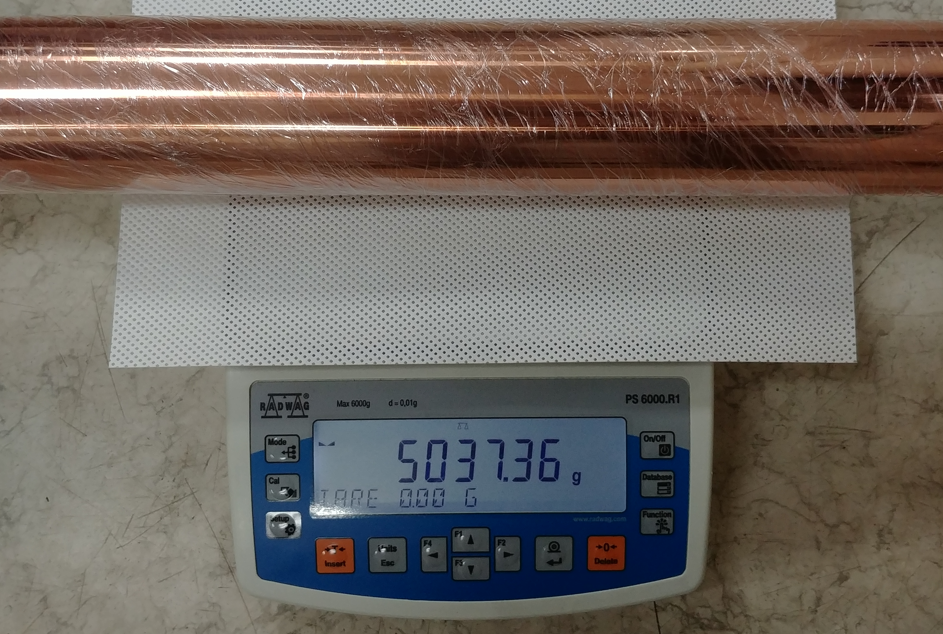}
  \caption{Tuning rod mass measured with an electronic scale.}
  \label{FIG:WT}
\end{figure}
In light of the cavity dimensions, it can be placed in the magnet
bore of the CAPP-12TB experiment~\cite{12TB_PRL,MAX}.
The pieces are all OFHC (oxygen free high conductivity) copper except
for an alumina axle of the tuning rod. In order to avoid crank arms
linking the axle and rod, our tuning rod has an off-axis axle
developed in Ref.~\cite{HAYSTAC_CAVITY}.
With this off-axis axle located 70.5~mm away from the cavity center, the
tuning rod sweeps a part of the cavity volume resulting in a frequency
range of about 0.99 to 1.19~GHz.
In order to extend the frequency tuning range, the axle can be shifted
47~mm toward the cavity center from the aforementioned axle location,
which extends it up to around 1.51~GHz. The total frequency tuning
range is about 0.52~GHz that corresponds to about 42\% with respect to
the central frequency of the tuning range.

\begin{figure}[h]
  \centering
  \includegraphics[width=0.9\textwidth]{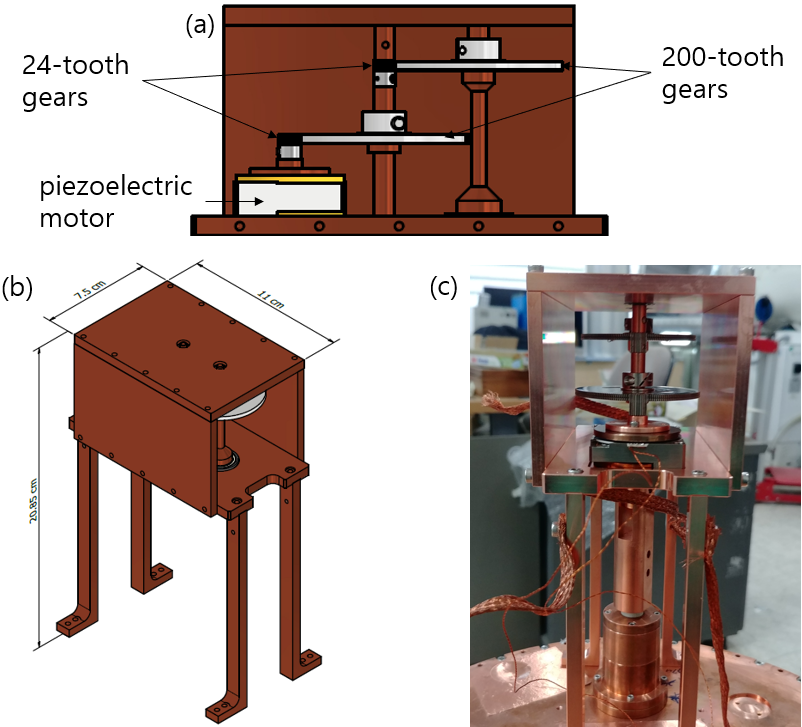}
  \caption{(a) shows the piezoelectric motor in combination with two
    200-tooth and two 24-tooth gears, (b) the overall structure and
    dimensions of the tuning mechanism driver. The tuning mechanism
    driver installed on the end cap of the cavity top is shown in
    (c).}
  \label{FIG:GEAR}
\end{figure}

The large tuning rod, which weighs about 5~kg as shown in
Fig.~\ref{FIG:WT}, was driven by a piezoelectric motor manufactured by
attocube~\cite{ATTOCUBE} in combination with gears.
Note that the tuning rod was manufactured with the minimum thickness
possible for the fine machining process to be carried out, not to
degrade the quality factor of the cavity system.
While such a heavy rod does have its advantages, such as the
capability to dampen environmental vibrations such as those coming
from vacuum pumps, it does require a tuning system powerful enough to
move the rod. This is where the gears become relevant.
The gear reduction ratio is about 69.4:1 resulting from the
combination of two 200-tooth and two 24-tooth plain gears as shown in
Fig.~\ref{FIG:GEAR}. The moment of inertia or rotational inertia of
the tuning rod considered in this work is about 32 times larger than
the tuning rod for the cavity used in our previous
work~\cite{12TB_PRL,MAX}, where the tuning mechanism driver was solely
a piezoelectric motor. Therefore, the gear reduction ratio of about
69.4 we chose in principle is a double of the necessary power and
would provide enough marginal power to handle the rotational inertia
by the tuning rod considered in this work unless other significant
issues appear.

The frequency range of 0.52~GHz could be realized by sweeping the
tuning rod from the cavity center to the cavity wall with an axle
located 47~mm away from the cavity center, but it is necessary to have
additional crank arms to link the axle and the rod.
Crank arms must be made of dielectric materials as opposed to
conductors to avoid unwanted capacitive effects, but are likely too
brittle to support our heavy tuning rod.
Furthermore, it increases the rotational inertia of the
tuning rod, hence an increase to the necessary driving power for the
frequency tuning mechanism.
As shown in Fig.~\ref{FIG:GEAR}(b), our tuning mechanism driver is
modularized, and therefore can easily adapt to the tuning axle's
location in the cavity. This is necessary since the axle's position
needs to be changed to access a different frequency range, by mounting
the tuning rod to a different set of holes in the end caps. It is,
however, nontrivial to switch the tuning axle itself as this requires
an additional cooling cycle to do so.

\section{Heating from the piezoelectric motor operation}\label{sec:piezo}
\begin{figure}[h]
  \centering
  \includegraphics[width=0.8\textwidth]{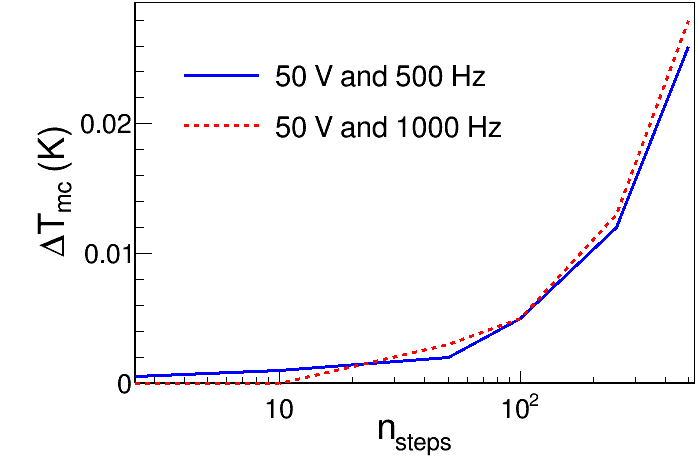}
  \caption{$\Delta T_{\rm mc}$ of $n_{\rm steps}$.
    The blue solid line and the red dashed lines denote measured
    $\Delta T_{\rm mc}(n_{\rm steps})$ with two different
    piezoelectric motor input driving frequencies of 500~Hz and
    1000~Hz, respectively, for a given piezoelectric motor input
    parameter $V_p=50$~V.}
  \label{FIG:DELTAT}
\end{figure}

The benefit from the frequency tuning mechanism employing the
piezoelectric motor is that the attocube piezoelectric motor ANR240
can be operated in cryogenic, high-magnetic-field, and high vacuum
environments~\cite{ATTOCUBE}, which is the typical axion haloscope
search environment.
On the other hand, the piezoelectric motor operation generates
unavoidable heat mostly from the power dissipated by the actuator.
This heat load is proportional to the number of the piezoelectric
motor steps $n_{\rm steps}$. With the gear combination developed in
this work, the required $n_{\rm steps}$ would increase in order to
move through frequencies at a reasonable pace such that the
experimental sensitivity is relatively preserved.
We measured the heat load on a dilution fridge LD400 manufactured by
Bluefors Oy whose guaranteed cooling power is 400~$\mu$W at
100~mK~\cite{BLUEFORS}.
Only a piezoelectric motor, without the cavity\footnote{Our Bluefors
dilution fridge cannot afford either the cavity dimension and
weight.}, was installed on the mixing chamber (mc) plate of the
Bluefors dilution fridge, establishing a good thermal link between the
two to mimic our axion dark matter search experimental conditions.
A time delay of 60~s was applied after the piezoelectric motor
operation, and afterwards we measured the temperature difference of
the mixing chamber $\Delta T_{\rm mc}$ depending on $n_{\rm steps}$
and the piezo driving frequency or signal repetition rate $f_d$ for a
given piezoelectric motor input voltage $V_p$ of 50~V. The base
temperature of the mixing chamber was about 40~mK. The blue solid line
and red dashed lines in Fig.~\ref{FIG:DELTAT} denote the measured
$\Delta T_{\rm mc}$ using $f_d$ of 500~Hz and 1000~Hz, respectively.
It is usually necessary to have such a time delay after the frequency
tuning to stabilize the system and depends on the target sensitivity
of the experiment, e.g., 30~s for the CAPP-12TB
experiment~\cite{12TB_PRL}, where $n_{\rm steps}$ was about 10 for a
frequency step of 10~kHz.
We also applied a 1~s delay for every fifth piezoelectric motor step,
e.g., 1~s delay in every 100~s for an $n_{\rm steps}$ of 500.
Assuming a good thermal link between the mixing chamber and cavity,
this $\Delta T_{\rm mc}$ is usually approximately the cavity physical
temperature increase $\Delta T_{\rm cavity}$.
The $\Delta T_{\rm cavity}$ subsequently increases the $T_n$, thus it
is crucial to prevent such heat load to maintain the experimental
sensitivity.
As shown in Fig.~\ref{FIG:DELTAT}, the temperature increase at the
Bluefors dilution fridge is about 10~mK due to the piezoelectric motor
operation with an $n_{\rm steps}$ of about 200 and about 30~mK when
$n_{\rm steps}$ is about 500, with the aforementioned
parameters. While the former is accepted for experiments depending on
their experimental sensitivity, the latter is not for the CAPP-12TB
sensitivity~\cite{12TB_PRL}.
Due to ongoing experiment, the DRS-1000 dilution fridge manufactured
by Leiden cryogenics BV~\cite{DRS-1000} was unavailable for use, and
we could not consider the case with the DRS-1000 dilution fridge whose
cooling power was measured to be 1~mW at 90~mK~\cite{12TB_PRL}.
Using the relation $\dot{Q}_{\rm mc}=84\dot{n}_3 T^2_{\rm mc}$ and the
aforementioned Bluefors specification, 400~$\mu$W at 100~mK, we could
expect three times stronger cooling power from the DRS-1000 fridge
compared to that from the Bluefors fridge, where $\dot{Q}_{\rm mc}$ is
the cooling power at the mixing chamber and $\dot{n}_3$ the $^3$He
molar circulation rate.
Hence, we could expect less temperature increase due to the
piezoelectric operation with the same parameters in the DRS-1000
dilution fridge.
Note that the temperature increase is practically independent of the
driving frequency of a piezoelectric motor, according to our
measurements shown in Fig.~\ref{FIG:DELTAT} for the given experimental
conditions.
\section{Validation of the cavity performance}\label{sec:cavityVal}
The cavity performance was first checked by measuring the cavity
unloaded quality factor of the cavity mode as a function of the cavity
mode frequency, $Q_0(\nu)$. $\Delta\nu$ depending on $V_p$, $f_d$, and
$n_{\rm steps}$ were then also measured for further validation of the
cavity performance. We performed these measurements on a 4-K
cryocooler due to a few aforementioned unavoidable reasons, but do not
expect significant drawbacks even in an $\mathcal{O}$(10 mK) environment
with the reasons described in the text below.
\subsection{$Q_0(\nu)$ measurements}\label{sec:cavityQ}
\begin{figure}[h]
  \centering
  \includegraphics[width=0.99\textwidth]{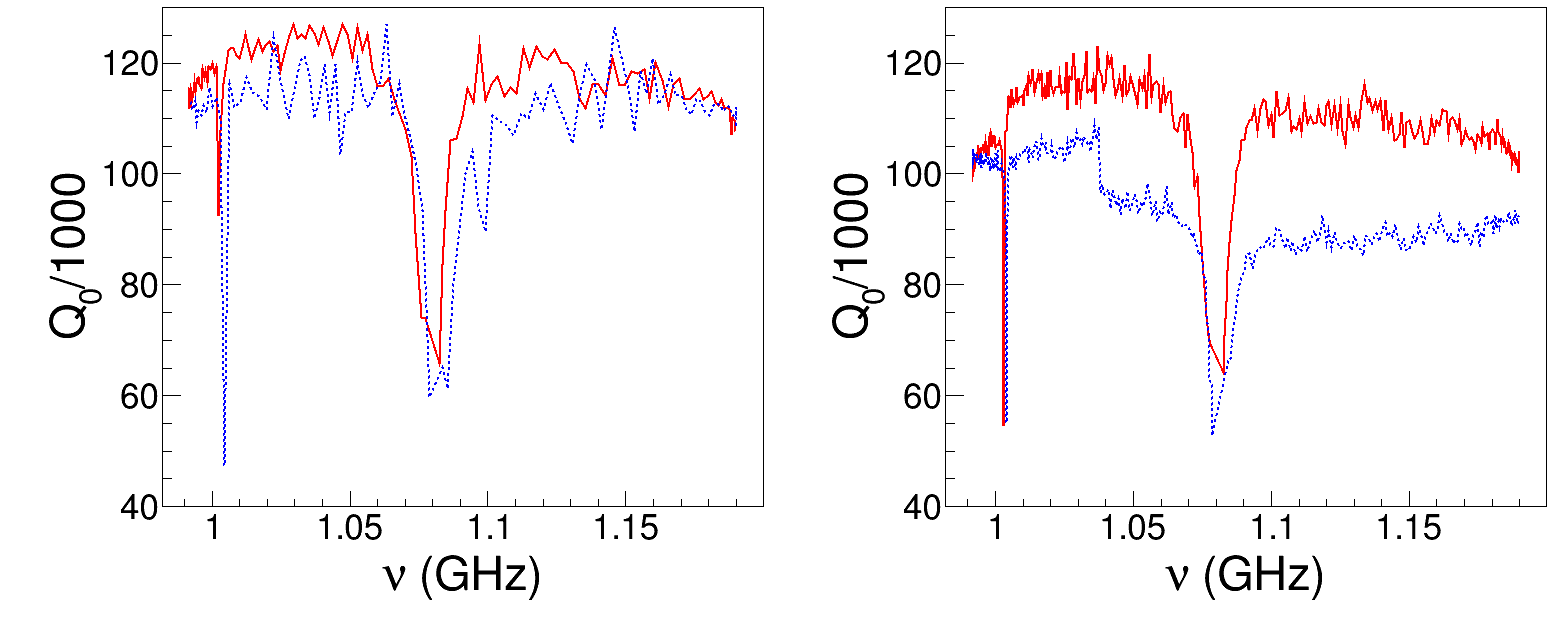}
  \caption{Left and right show the $Q_0(\nu)$ measurements with
    $V_p=60$~V and $f_d=1500$~Hz and those of 50~V and 1000~Hz,
    respectively. The red solid lines and blue dots were obtained by
    the clockwise tuning rod rotation from 0$^{\circ}$ to
    180$^{\circ}$ and from 180$^{\circ}$ to 360$^{\circ}$,
    respectively, where the tuning rod rotation angle of 0$^{\circ}$
    corresponds to Fig.~\ref{FIG:CAVITY}(b).
    Due to the larger frequency tuning step with the larger driving
    power, the left data was measured relatively less times compared
    to the right data. The calibration using network analyzer
    calibration kits was applied for the left data, but not for the
    right data. The two mode crossings around 1~GHz and 1.08~GHz are
    also identified from cavity simulations~\cite{CST,COMSOL}. The
    latter is actually an “avoided crossing”~\cite{HAYSTAC_CAVITY},
    making the dip in $Q_0$ wider.}  
  \label{FIG:Q0}
\end{figure}

\begin{figure*}[h]
  \centering
  \subfigure{\includegraphics[width=0.55\textwidth]{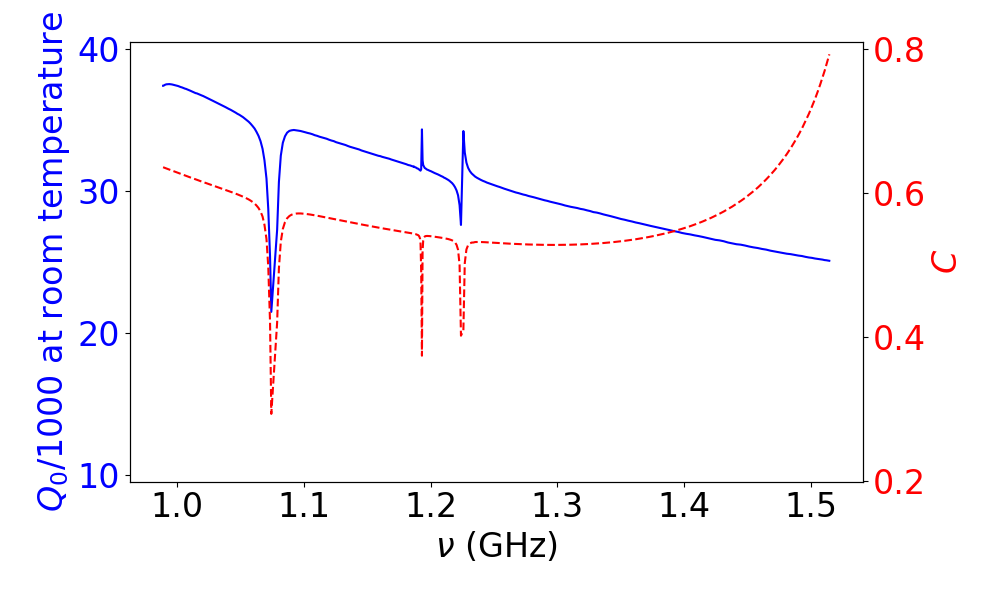}}
  \subfigure{\includegraphics[width=0.43\textwidth]{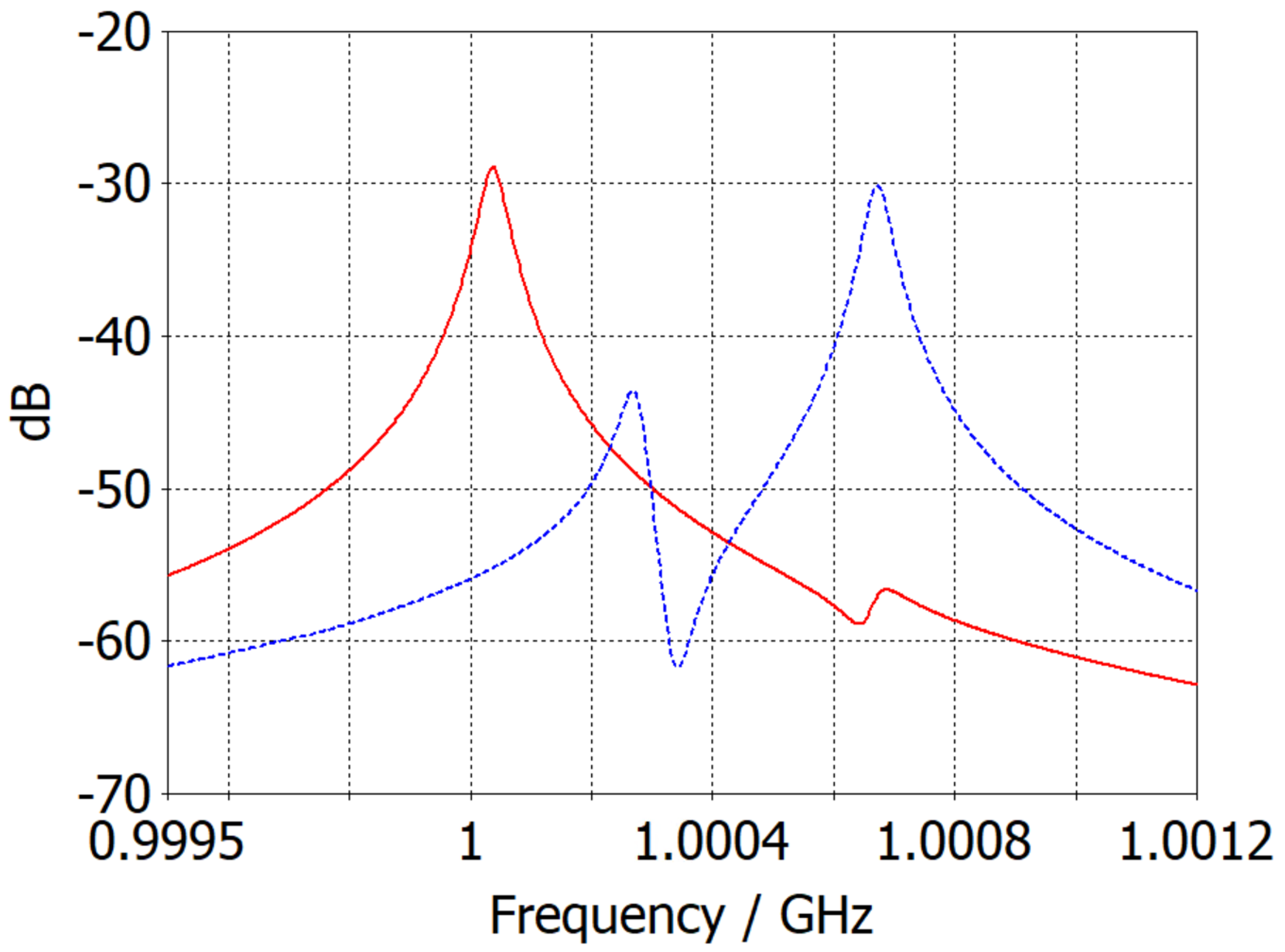}}
  \caption{Left shows the $Q_0$ (blue solid line) and $C$ (red
    dashed-line) of the TM$_{010}$ modes as a function of $\nu$ from
    the eigenmode simulation, where $C$s were calculated from a
    constant magnetic field.    
    Right was obtained from the frequency domain simulation and shows
    the transmission at two different resonant frequencies displaying
    the mode crossing around 1~GHz, where the larger peaks in the red
    solid line and blue dashed-line are the signal modes and the
    smaller ones in those are the intruder modes, respectively.}
  \label{FIG:SIMQC}
\end{figure*}

The $Q_L(\nu)$ was read by the transmission from a weakly coupled
antenna to a strongly coupled antenna and the measurements were
performed over the full rotation of the tuning rod.
We measured $Q_L$ without averaging to proceed quickly through the
tuning range, which resulted in the $Q_L$ fluctuations propagating to
the $Q_0$ fluctuations appearing in Fig.~\ref{FIG:Q0}.
In order to get $Q_0$ from $Q_L$, we also measured the reflection of
the antenna strongly coupled to the cavity.
Following the procedure using the Smith chart
data~\cite{8TB_PRL,8TB_NIM}, we obtained the coupling coefficient
$\beta$. The $Q_0$ was then calculated from the relation
$Q_0=(1+\beta)Q_L$. The results are shown in Fig.~\ref{FIG:Q0} which
were obtained from the full clockwise rotation of the tuning rod.
The results from driving in the opposite direction were similar and
left out.
As shown in Fig.~\ref{FIG:Q0}, our tunable cavity developed in this work
shows not only good quality factors over the full range of the
resonance frequency, but also the capability of the full rotation of
the tuning rod by the tuning mechanism driver with the piezoelectric
motor combination with gears.
Even though we did not manage to perform further measurements by switching
the tuning axle, it would be done without too much trouble as
aforementioned with the modularized tuning driver.
Here, instead, simulation results corresponding to the frequency range
up to around 1.51~GHz are given in Fig.~\ref{FIG:SIMQC}.
The two modes turn out to cross with little mixing according to
simulations, hence the eigenmode simulation results shown in the left
plot of Fig.~\ref{FIG:SIMQC} do not catch the drop-off of $Q_0$ around
1~GHz, which was observed in data shown in Fig.~\ref{FIG:Q0}. Such a
mode crossing with little mixing was also reported in
Ref.~\cite{ACTION}. The mode crossing around 1~GHz was identified by
the frequency domain simulation as shown in the right plot of
Fig.~\ref{FIG:SIMQC}.

We noted that the transmission line calibration between the
network analyzer (NA) and cavity using the NA calibration kits is
crucial to measure the coupling coefficient $\beta$, and subsequently
$Q_0$ as shown in Fig.~\ref{FIG:Q0}, where the calibration was applied
to the left, but not to the right. Since the cavity is approximately
symmetric in the azimuthal coordinates, without any external coupling
effects $Q_0$ should be also be symmetric in those coordinates as
shown in the left plot of Fig.~\ref{FIG:Q0}.
The asymmetric $Q_0$ shown in the right plot of Fig.~\ref{FIG:Q0} can
be understood as the following.
In terms of an $xy$ plane view with the cavity center as the origin,
the two antenna ports are located in the $x$-axis, where the weakly
coupled port is in positive $x$ and the strongly coupled port in
negative $x$. Since the tuning axle is located in negative $y$, the
tuning rod location with respect to the strongly coupled port is not
symmetric when it rotates.
The red solid lines in Fig.~\ref{FIG:Q0} were measured when the tuning
rod is close to the strong port, where the electric field was pushed
by our conductor tuning rod, and thus the field is located away from
the strong port resulting in a smaller $\beta$. The blue dots in
Fig.~\ref{FIG:Q0} are opposite cases to the red solid lines, and
become overcoupled with larger $\beta$ values. According to our
measurements with and without applying the NA calibration kit shown in
Fig.~\ref{FIG:Q0}, overcoupling was not detected properly without
applying the NA calibration kit.

Although the $Q_0(\nu)$ measurements were done in temperatures of
about 4~K for the cavity,
their values would not change even in an $\mathcal{O}$(10 mK)
environment due to the anomalous skin effect in metals~\cite{PIPPARD},
i.e., copper for our cavity.
\subsection{$\Delta\nu$ measurements}\label{sec:deltf}
\begin{figure}[h]
  \centering
  \includegraphics[width=0.8\textwidth]{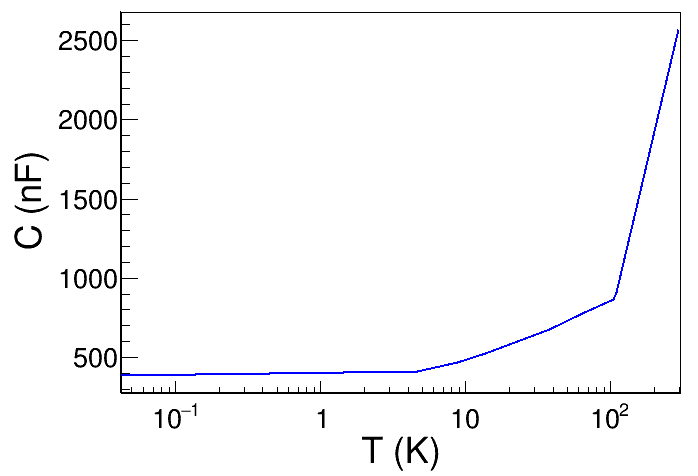}
  \caption{Measurements for the $C(T)$ of the piezoelectric motor
    system from room temperature to 40 mK on the dilution fridge.}
  \label{FIG:COFT}
\end{figure}

These $\Delta\nu$ measurements were also performed in temperatures of
about 4~K, and therefore the stored power in the piezoelectric motor
system is expected to be degraded in an $\mathcal{O}$(10~mK) environment
due to the temperature dependence of capacitance. First we measured
the temperature dependent capacitance $C(T)$ to see how much power
degradation is expected when it operates in an $\mathcal{O}$(10~mK)
environment.
The $C(T)$ can be measured directly from the attocube systems' piezo
motor controller ANC350~\cite{ATTOCUBE} and was measured down to a
temperature of about 40~mK on the aforementioned Bluefors dilution
fridge LD400. Figure~\ref{FIG:COFT} shows the $C(T)$ measurement
resulting in $\frac{C(40~\rm{mK})}{C(4~\rm{K})}\sim0.93$, where the
factor 0.93 is also true for the piezoelectric motor's driving power
for a given $V_p$. As mentioned in Sec.~\ref{sec:cavity}, the
chosen gear reduction ratio is a double of the necessary power, so we
do not expect significant effects from the piezo power degradation in
an $\mathcal{O}$(10 mK) environment.

For the $\Delta\nu$ measurements, we limited the tuning range from
1.09 to 1.10~GHz to compare the cavity system here with the one used
in the CAPP-12TB experiment~\cite{12TB_PRL}. The $\Delta\nu$ of the
CAPP-12TB experiment was about 10~kHz resulting from an
$n_{\rm steps}$ of about 10, $V_p=50$~V, and $f_d=1000$~Hz. The
frequency tuning with the tuning mechanism parameters did not
introduce significant heat load to the system, which was one of the
key ingredients that resulted in the success of the CAPP-12TB
experiment. Note that this is not a like-for-like comparison due to
the different cavity regions swept by different tuning mechanisms, but
to check if the experimental sensitivity in this work is comparable to
that of CAPP-12TB reflecting every experimental aspect as much as
possible.

\begin{figure*}[h]
  \centering
  \subfigure{\includegraphics[width=0.49\textwidth]{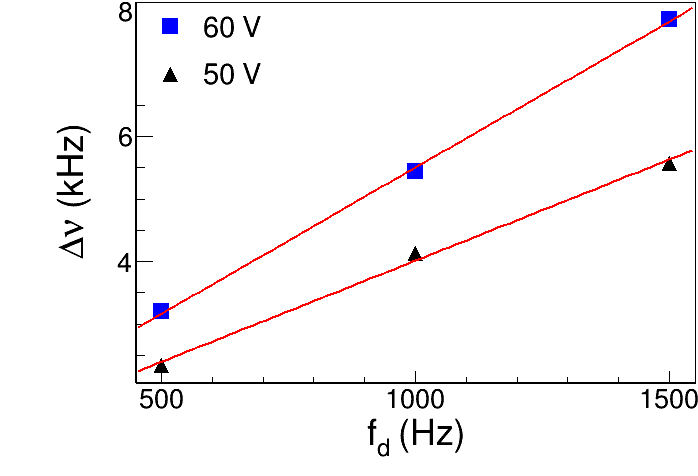}}
  \subfigure{\includegraphics[width=0.49\textwidth]{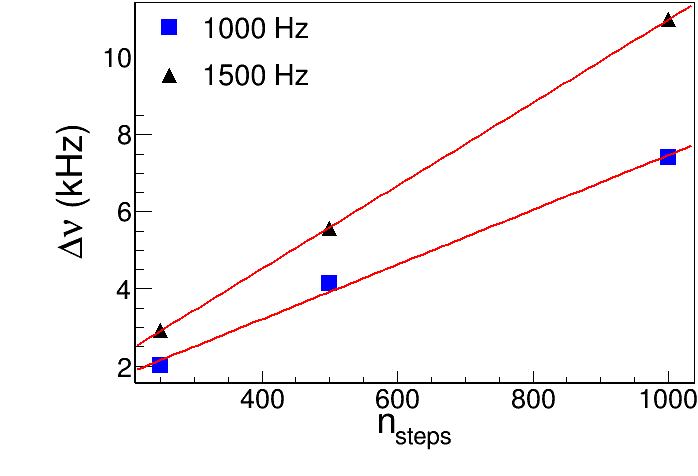}}
  \caption{Left shows the $\Delta\nu$ measurements as a function of
    $f_d$ for an $n_{\rm steps}$ of 500, where the blue rectangles and
    the black triangles were measured with $V_p$ of 60~V and 50~V,
    respectively.    
    Right shows the $\Delta\nu$ measurements as a function of
    $n_{\rm steps}$ with a $V_p$ of 50~V, where the blue rectangles
    and the black triangles were measured with $f_d$ of 1000~Hz and
    1500~Hz, respectively.    
    The red lines are the fits described in the text.}
  \label{FIG:DELF}
\end{figure*}

The left plot of Fig.~\ref{FIG:DELF} shows the $\Delta\nu(f_d)$ for an
$n_{\rm steps}$ of 500, where the blue rectangles and the black
triangles were measured with $V_p$ of 60~V and 50~V, respectively.
Linear functions $\Delta\nu(f_d)=0.83+0.0047f_d$ and
$\Delta\nu(f_d)=0.78+0.0032f_d$ were obtained from the linear fits for
the blue rectangles and black triangles, respectively.
Therefore, the $\Delta\nu$ is about 4~kHz from the tuning mechanism
conditions, $n_{\rm steps}=500$, $V_p=50$~V, and $f_d=1000$~Hz,
whose conditions could increase the $T_{\rm mc}$ of about 30~mK
already for a dilution fridge system according to our measurements
shown in Fig.~\ref{FIG:DELTAT}.
In order to utilize the tuning mechanism developed here, we can employ
a $\Delta\nu$ of 2~kHz and approximately a fifth of
$\Delta t_{\rm 10~kHz}$ per each tuning step, where $\Delta t$ is an
integration time in haloscope search experiments and
$\Delta t_{\rm 10~kHz}$ is that for the case with a $\Delta\nu$ of
10~kHz.
This relatively finer tuning step approach that has been employed by
the Axion Dark Matter eXperiment~\cite{ADMX1,ADMX2,ADMX3,ADMX4} does
not degrade the statistical sensitivity, but would elongate the scanning rate
depending on the aforementioned time delay for the system
stabilization, mainly cooling. In order to find the $n_{\rm steps}$
for the finer tuning step of 2~kHz, we measured
$\Delta\nu(n_{\rm steps})$ with $V_p=50$~V which is shown as the blue
rectangles in the right plot of Fig.~\ref{FIG:DELF}. From the linear
fit function of the rectangles
$\Delta\nu(n_{\rm steps})=0.374+0.0071n_{\rm steps}$,
we found that $n_{\rm steps}=230$ results in $\Delta\nu\sim2$~kHz.
Since we found no significant temperature increase with a higher $f_d$
as shown in Fig.~\ref{FIG:DELTAT}, we also measured them with a higher
$f_d$ of 1500~Hz which is shown as the black triangles in the same
plot. From the linear fit function of the triangles
$\Delta\nu(n_{\rm steps})=0.213+0.01081n_{\rm steps}$, we found that
$n_{\rm steps}=170$ results in $\Delta\nu\sim2$~kHz. Taking into
consideration the results shown in Fig.~\ref{FIG:DELTAT}, the $\Delta
T_{\rm mc}$ from the piezoelectric motor operation with
$n_{\rm steps}=170$, $V_p=50$~V, and $f_d=1500$~Hz, would be less than
10~mK at an axion dark matter experiment employing the Bluefors
LD400. One can na\"ively expect further suppression of
$\Delta T_{\rm mc}$ at the experiment employing the Leiden DRS-1000
thanks to its stronger cooling power, and the cooling time is also
expected to be shorter.
With three times stronger cooling power we can assume a time delay of
20~s, which changes the total running time to move 10~kHz to
600~s. This is 13\% longer than the CAPP-12TB case~\cite{12TB_PRL},
but is still in a generally acceptable range.
\section{Summary}\label{sec:summary}
We developed an enhanced tunable cavity system for axion dark matter
search experiments. Our cavity system employed a large and accordingly
heavy tuning rod to increase the frequency tuning range. With a
piezoelectric motor in combination with gears, we were able to drive a
heavy tuning rod and realized a wideband tunable cavity whose
frequency range is about 42\% of the central frequency of the tuning
range.
By employing a relatively finer tuning step of 2~kHz, we expect the
aforementioned delay of 13\% in the scanning rate, thus insignificant
experimental sensitivity drop-off of about 6\% even compared with the
experimental sensitivity coming from the best scanning rate to
date~\cite{12TB_PRL}.
This first tuning mechanism driver with a piezoelectric motor in
combination with gears can drive much bigger power than that with the
piezoelectric motor only. Our approach therefore can be useful to
axion dark matter search experiments requiring heavy tuning mechanisms
with several conductor tuning rods toward higher frequencies or a
large dielectric chunk of tuning rod toward lower frequencies, and
also to drive heavy loads under extreme environments.

\acknowledgments
This work was supported by the Institute for Basic Science (IBS) under
Project Code No. IBS-R017-D1-2024-a00 and a Korea University
Grant. B. R. Ko acknowledges G. Rybka suggested the tuning mechanism
driver idea.

\end{document}